\documentclass[aps,showpacs,twocolumn]{revtex4}

\usepackage{graphicx}
\usepackage{epstopdf}
\usepackage{dcolumn}

\newcommand{\ba}{\begin{eqnarray}}
\newcommand{\ea}{\end{eqnarray}}

\begin{document}
\title{Partial dynamical symmetry
in quantum Hamiltonians with higher-order terms}

\author{J.E.~Garc\'\i a-Ramos$^1$,
A.~Leviatan$^2$,
and P.~Van Isacker$^3$}
\affiliation{
$^1$Department of Applied Physics, University of Huelva,
21071 Huelva, Spain}
\affiliation{
$^2$Racah Institute of Physics,
The Hebrew University, Jerusalem 91904, Israel}
\affiliation{
$^3$Grand Acc\'el\'erateur National d'Ions Lourds,
CEA/DSM--CNRS/IN2P3, B.P.~55027, F-14076 Caen Cedex 5, France}
\date{\today}

\begin{abstract}
A generic procedure is proposed
to construct many-body quantum Hamiltonians
with partial dynamical symmetry.
It is based on a tensor decomposition of the Hamiltonian
and allows the construction of a hierarchy of interactions
that have selected classes of solvable states.
The method is illustrated in the SO(6) limit
of the interacting boson model of atomic nuclei
and applied to the nucleus $^{196}$Pt.
\end{abstract}

\maketitle

The concept of dynamical symmetry (DS)
is now widely accepted to be of central importance
in our understanding of many-body systems.
In particular, it had a major impact
on developments in nuclear~\cite{Iachello87} and
molecular~\cite{Iachello95} physics
and made significant contributions
to virtually all areas of many-body physics~\cite{Bohm88}.
Its basic paradigm is to write the Hamiltonian
of the system under consideration
in terms of Casimir operators
of a set of nested algebras.
Its hallmarks are
(i) solvability of the complete spectrum,
(ii) existence of exact quantum numbers for all eigenstates,
and (iii) pre-determined symmetry-based structure of the eigenfunctions,
independent of the Hamiltonian's parameters.

The merits of a DS are self-evident. However, in most applications to
realistic systems, the predictions of an exact DS are rarely fulfilled
and one is compelled to break it.
More often one finds that the assumed symmetry is
not obeyed uniformly, {\it i.e.}, is fulfilled by only some states but
not by others. The need to address such situations has led to the
introduction of
partial dynamical symmetries (PDSs).
The essential idea is to relax the stringent conditions
of {\em complete} solvability
so that the properties (i)--(iii)
are only partially satisfied.
Partiality comes in three different guises:
(a)~part of the eigenspectrum retains
all the DS quantum numbers~\cite{Alhassid92,Leviatan96},
(b)~the entire eigenspectrum retains
part of the DS quantum numbers~\cite{Leviatan86,Isacker99},
and (c)~part of the eigenspectrum retains
part of the DS quantum numbers~\cite{Leviatan02}.
PDS of various types have been shown to be relevant to
nuclear~\cite{Leviatan96,Leviatan86,Isacker99,Leviatan02,Escher00}
and molecular~\cite{Ping97} spectroscopy, to systems with mixed chaotic and
regular dynamics~\cite{WAL93} and to
quantum phase transitions~\cite{lev07}.

We emphasize that DS as well as its generalization PDS
are notions that are not restricted to a specific model
but can be applied to any quantal system of interacting particles,
bosons and fermions.
In this Letter we propose a generic method
to construct quantum Hamiltonians
with PDS of type~(a)
and show that its existence is closely related
to the order of the interaction among the particles.
The procedure is discussed in general terms
and subsequently illustrated with a nuclear-physics example
with an application to the nucleus $^{196}$Pt.

The analysis starts from the chain of nested algebras
\begin{equation}
\begin{array}{ccccccc}
G_{\rm dyn}&\supset&G&\supset&\cdots&\supset&G_{\rm sym}\\
\downarrow&&\downarrow&&&&\downarrow\\[0mm]
[h]&&\langle\Sigma\rangle&&&&\Lambda
\end{array}
\label{chain}
\end{equation}
where, below each algebra,
its associated labels of irreducible representations (irreps) 
are given. Eq.~(\ref{chain}) implies 
that $G_{\rm dyn}$ is the dynamical algebra of the system 
such that operators of all physical observables 
can be written in terms of its generators~\cite{Frank94}; 
a single irrep of $G_{\rm dyn}$
contains all states of relevance in the problem.
In contrast, $G_{\rm sym}$ is the symmetry algebra
and a single of its irreps contains states that are degenerate in energy.
A frequently encountered example is $G_{\rm sym}={\rm SO}(3)$,
the algebra of rotations in 3 dimensions,
with its associated quantum number
of total angular momentum $L$.
Other examples of conserved quantum numbers
can be the total spin $S$ in atoms
or total isospin $T$ in atomic nuclei.

The classification~(\ref{chain}) is generally valid
and does not require conservation of particle number.
Although the extension from DS to PDS
can be formulated under such general conditions,
let us for simplicity of notation assume in the following
that particle number is conserved.
All states, and hence the representation $[h]$,
can then be assigned a definite particle number $N$.
For $N$ identical particles
the representation $[h]$
of the dynamical algebra $G_{\rm dyn}$
is either symmetric $[N]$ (bosons)
or antisymmetric $[1^N]$ (fermions)
and will be denoted, in both cases, as $[h_N]$.
For particles that are non-identical
under a given dynamical algebra $G_{\rm dyn}$,
a larger algebra can be chosen
such that they become identical under this larger algebra
(generalized Pauli principle)~\cite{Note1}.
The occurrence of a DS of the type~(\ref{chain})
signifies that eigenstates can be labeled as
$|[h_N]\langle\Sigma\rangle\dots\Lambda\rangle$;
additional labels (indicated by $\dots$)
are suppressed in the following.
Likewise, operators can be classified
according to their tensor character under~(\ref{chain})
as $\hat T_{[h_n]\langle\sigma\rangle\lambda}$.
Of specific interest in the construction of a PDS
associated with the reduction~(\ref{chain}),
are the $n$-particle annihilation operators $\hat T$
which satisfy the property
\begin{equation}
\hat T_{[h_n]\langle\sigma\rangle\lambda}
|[h_N]\langle\Sigma_0\rangle\Lambda\rangle=0,
\label{anni}
\end{equation}
for all possible values of $\Lambda$
contained in a given irrep~$\langle\Sigma_0\rangle$.
Any number-conserving normal-ordered interaction
written in terms of these annihilation operators
(and their Hermitian conjugates which transform as the
corresponding conjugate irreps)
can be added to the Hamiltonian with a DS~(\ref{chain}),
while still preserving the solvability
of states with $\langle\Sigma\rangle=\langle\Sigma_0\rangle$.
The annihilation condition~(\ref{anni}) is satisfied
if none of the $G$ irreps $\langle\Sigma\rangle$
contained in the $G_{\rm dyn}$ irrep $[h_{N-n}]$
belongs to the $G$ Kronecker product
$\langle\Sigma_0\rangle\times\langle\sigma\rangle$.
So the problem of finding interactions
that preserve solvability
for part of the states~(\ref{chain})
is reduced to carrying out a Kronecker product.

Let us now illustrate this procedure
with an example taken from nuclear physics
where the concept of DS
has been applied successfully
in the context of the interacting boson model (IBM)~\cite{Iachello87}.
In this model, low-energy collective states
are described in terms of $N$ interacting monopole $(s)$ and
quadrupole $(d)$ bosons representing valence nucleon pairs.
The dynamical algebra is $G_{\rm dyn}={\rm U}(6)$
and the symmetry algebra is $G_{\rm sym}={\rm SO}(3)$.
Three DS limits occur in the model
with leading subalgebras U(5), SU(3), and SO(6),
corresponding to typical collective spectra observed in nuclei,
vibrational, rotational, and $\gamma$-unstable, respectively.
Here we focus on the SO(6) limit with predictions that were found to
correspond closely to the empirical structure of some platinum
nuclei~\cite{Cizewski78} as well as of other nuclei, notably,
around mass number $A=130$~\cite{Casten85}.

The classification of states
in the SO(6) limit of the IBM is
\begin{equation}
\begin{array}{ccccccccc}
{\rm U}(6)&\supset&{\rm SO}(6)&\supset&{\rm SO}(5)&
\supset&{\rm SO}(3)&\supset&{\rm SO}(2)\\
\downarrow&&\downarrow&&\downarrow&&\downarrow&&
\downarrow\\[0mm]
[N]&&\langle\Sigma\rangle&&(\tau)&\nu_\Delta& L&&M
\end{array}.
\label{chainso6}
\end{equation}
The multiplicity label $\nu_\Delta$ in the
${\rm SO(5)}\supset {\rm SO(3)}$ reduction
will be omitted in the following when it is not needed.
The eigenstates $|[N]\langle\Sigma\rangle(\tau)\nu_\Delta LM\rangle$
are obtained with a Hamiltonian
with SO(6) DS which, for one- and two-body interactions, can be
transcribed in the form
\ba
\hat{H}_{\rm DS} &=& A\,\hat{P}_{+}\hat{P}_{-} +
B\, \hat{C}_{{\rm SO(5)}} + C\,\hat{C}_{{\rm SO(3)}}.
\label{hDS}
\ea
Here $\hat{C}_{G}$ denotes the quadratic Casimir operator of $G$,
$\hat{P}_{+}\equiv{1\over2}(s^\dag s^\dag-d^\dag\cdot d^\dag)$,
$4\hat{P}_{+}\hat{P}_{-} = \hat{N}(\hat{N}+4)-\hat C_{{\rm SO(6)}}$
and $\hat{P}_{-}= \hat{P}_{+}^{\dagger}$.
The total boson number operator,
$\hat{N}=\hat{n}_s + \hat{n}_d$, is the linear Casimir of
U(6) and is a constant for all $N$-boson states.
The spectrum of $\hat{H}_{\rm DS}$ is completely solvable with eigenenergies
\ba
E_{\rm DS} &=& {\textstyle{\frac{1}{4}}}
A\,(N-\Sigma)(N+\Sigma + 4) + B\,\tau(\tau+3)
\nonumber\\
&&
+\, C\,L(L+1).
\label{eDS}
\ea

In contrast,
Hamiltonians with SO(6) PDS
preserve the analyticity of only
a {\em subset}
of the states~(\ref{chainso6}).
The construction of interactions with this property
requires $n$-boson creation and annihilation operators
with definite tensor character in the basis~(\ref{chainso6}):
\begin{equation}
\hat B^\dag_{[n]\langle\sigma\rangle(\tau)lm},
\;\;
\tilde{B}_{[n^5]\langle\sigma\rangle(\tau)lm}
\equiv
(-1)^{l-m}
\left(\hat B^\dag_{[n]\langle\sigma\rangle(\tau)l,-m}\right)^\dag.
\label{tenso6}
\end{equation}
Of particular interest are tensor operators with $\sigma<n$.
They have the property
\begin{equation}
\tilde{B}_{[n^5]\langle\sigma\rangle(\tau)lm}
|[N]\langle N\rangle(\tau)\nu_\Delta LM\rangle=0,
\qquad
\sigma<n,
\label{anniso6}
\end{equation}
for all possible values of $\tau,L$ contained in the
SO(6) irrep $\langle N\rangle$.
This is so because the action of
$\tilde{B}_{[n^5]\langle\sigma\rangle(\tau)lm}$
leads to an $(N-n)$-boson state
that contains the SO(6) irreps
$\langle\Sigma\rangle=\langle N-n-2i\rangle,\,i=0,1,\dots$
which cannot be coupled with $\langle\sigma\rangle$
to yield $\langle\Sigma\rangle=\langle N\rangle$, since $\sigma<n$.
Number-conserving normal-ordered interactions that are constructed out
of such tensors with $\sigma<n$
(and their Hermitian conjugates)
thus have $|[N]\langle N\rangle(\tau)\nu_\Delta LM\rangle$
as eigenstates with zero eigenvalue.

A systematic enumeration of all interactions with this property
is a simple matter of SO(6) coupling.
For {\it one-body} operators one has
\begin{equation}
\hat B^\dag_{[1]\langle1\rangle(0)00}=
s^\dag\equiv b_0^\dag,
\quad
\hat B^\dag_{[1]\langle1\rangle(1)2m}=
d_m^\dag\equiv b_{2m}^\dag,
\label{one}
\end{equation}
and no annihilation operator has the
property~(\ref{anniso6}).

Coupled {\it two-body} operators are of the form
\begin{equation}
\hat B^\dag_{[2]\langle\sigma\rangle(\tau)lm}\propto
\sum_{\tau_k\tau_{k'}}\sum_{kk'}
C^{\langle\sigma\rangle(\tau)l}_{\langle 1\rangle(\tau_k)k,
\langle 1\rangle(\tau_{k'})k'}
(b^\dag_k\times b_{k'}^\dag)^{(l)}_m,
\label{two}
\end{equation}
where $(\cdot\times\cdot)^{(l)}_m$
denotes coupling to angular momentum $l$
and the $C$-coefficients
are known ${\rm SO}(6)\supset{\rm SO}(5)\supset{\rm SO}(3)$
isoscalar factors~\cite{Frank94}.
This leads to the normalized two-boson SO(6) tensors
shown in Table~\ref{tensors}.
\begin{table}
\caption{\label{tensors}
Normalized two- and three-boson SO(6) tensors.}
\begin{ruledtabular}
\begin{tabular}{ccccl}
$n$&$\sigma$&$\tau$&$l$&
$\hat B^\dag_{[n]\langle\sigma\rangle(\tau)lm}$\\[4pt]
\hline
2& 2& 2& 4& $\sqrt{1\over2}(d^\dag\!\times\!d^\dag)^{(4)}_m$\\[2pt]
2& 2& 2& 2& $\sqrt{1\over2}(d^\dag\!\times\!d^\dag)^{(2)}_m$\\[2pt]
2& 2& 1& 2& $(s^\dag\!\times\!d^\dag)^{(2)}_m$\\[2pt]
2& 2& 0& 0& $\sqrt{5\over{12}}(s^\dag\!\times\!s^\dag)^{(0)}_0
            +\sqrt{1\over{12}}(d^\dag\!\times\!d^\dag)^{(0)}_0$\\[2pt]
2& 0& 0& 0& $\sqrt{1\over{12}}(s^\dag\!\times\!s^\dag)^{(0)}_0
            -\sqrt{5\over{12}}(d^\dag\!\times\!d^\dag)^{(0)}_0$\\[2pt]
3& 3& 3& 6& $\sqrt{1\over6}((d^\dag\!\times\!d^\dag)^{(4)}
             \!\times\!d^\dag)^{(6)}_m$\\[2pt]
3& 3& 3& 4& $\sqrt{7\over{22}}((d^\dag\!\times\!d^\dag)^{(2)}
             \!\times\!d^\dag)^{(4)}_m$\\[2pt]
3& 3& 3& 3& $\sqrt{7\over{30}}((d^\dag\!\times\!d^\dag)^{(2)}
             \!\times\!d^\dag)^{(3)}_m$\\[2pt]
3& 3& 3& 0& $\sqrt{1\over6}((d^\dag\!\times\!d^\dag)^{(2)}
             \!\times\!d^\dag)^{(0)}_0$\\[2pt]
3& 3& 2& 4& $\sqrt{1\over2}((s^\dag\!\times\!d^\dag)^{(2)}
             \!\times\!d^\dag)^{(4)}_m$\\[2pt]
3& 3& 2& 2& $\sqrt{1\over2}((s^\dag\!\times\!d^\dag)^{(2)}
             \!\times\!d^\dag)^{(2)}_m$\\[2pt]
3& 3& 1& 2& $\sqrt{7\over 16}((s^\dag\!\times\!s^\dag)^{(0)}
             \!\times\!d^\dag)^{(2)}_m
             +\sqrt{{5}\over{112}}((d^\dag\!\times\!d^\dag)^{(0)}
             \!\times\!d^\dag)^{(2)}_m$\\[2pt]
3& 3& 0& 0& $\sqrt{5\over{48}}((s^\dag\!\times\!s^\dag)^{(0)}
             \!\times\!s^\dag)^{(0)}_0
            +\sqrt{3\over{16}}((s^\dag\!\times\!d^\dag)^{(2)}
             \!\times\!d^\dag)^{(0)}_0$\\[2pt]
3& 1& 1& 2& $\sqrt{1\over{16}}((s^\dag\!\times\!s^\dag)^{(0)}
             \!\times\!d^\dag)^{(2)}_m
            -\sqrt{5\over{16}}((d^\dag\!\times\!d^\dag)^{(0)}
             \!\times\!d^\dag)^{(2)}_m$\\[2pt]
3& 1& 0& 0& $\sqrt{1\over{16}}((s^\dag\!\times\!s^\dag)^{(0)}
             \!\times\!s^\dag)^{(0)}_0
            -\sqrt{5\over{16}}((s^\dag\!\times\!d^\dag)^{(2)}
             \!\times\!d^\dag)^{(0)}_0$\\
\end{tabular}
\end{ruledtabular}
\end{table}
There is one operator with $\sigma<n=2$
and it gives rise to the following SO(6)-invariant interaction
\begin{eqnarray}
&&\hat B^\dag_{[2]\langle0\rangle(0)00}
\tilde{B}_{[2^5]\langle0\rangle(0)00}
={\textstyle{1\over3}}\hat P_+\hat P_-,
\end{eqnarray}
which is simply the SO(6) term in $\hat{H}_{\rm DS}$, Eq.~(\ref{hDS}).
This proves that a two-body interaction
which is diagonal in
$|[N]\langle N\rangle(\tau)\nu_\Delta LM\rangle$
is diagonal in all states
$|[N]\langle\Sigma\rangle(\tau)\nu_\Delta LM\rangle$.
This result is valid in the SO(6) limit of the IBM, but not in general.
For example, from a tensor decomposition
of two-boson operators in SU(3)
one concludes that the SU(3) limit of the IBM
{\em does} allow a PDS with two-body
interactions~\cite{Alhassid92,Leviatan96}.

{\it Three-body} operators with good SO(6) labels
can be obtained from an expansion similar to~(\ref{two})
and this leads to the normalized three-boson SO(6) tensors
shown in Table~\ref{tensors}.
In terms of the $\hat{P}_{+}$ operator introduced above,
the two operators with $\sigma<n=3$ are
\begin{equation}
\hat B^\dag_{[3]\langle1\rangle(1)2m}=
{\textstyle{1\over2}}\hat P_+d^\dag_m,
\quad
\hat B^\dag_{[3]\langle1\rangle(0)00}=
{\textstyle{1\over2}}\hat P_+s^\dag,
\label{three}
\end{equation}
and from these one can construct
the interactions with an SO(6) PDS.
The only three-body interactions that are partially solvable in SO(6)
are thus $\hat P_+\hat n_s\hat P_-$
and $\hat P_+\hat n_d\hat P_-$, involving the $s$-
and $d$-boson number operators.
Since the combination $\hat P_+(\hat n_s+\hat n_d)\hat P_-
= (\hat{N} -2)\hat{P}_{+}\hat{P}_{-}$
is completely solvable in SO(6),
there is only one genuine partially solvable three-body interaction
which can be chosen as $\hat P_+\hat n_s\hat P_-$,
with tensorial components $\sigma=0,2$.

The generalization to higher orders now suggests itself.
For example, {\it four-body} interactions with SO(6) PDS
are written in terms of
$\hat B^\dag_{[4]\langle2\rangle(\tau)lm}$
and $\hat B^\dag_{[4]\langle0\rangle(0)00}$,
and Hermitian conjugate operators.
Without loss of generality,
these operators can be written as
\begin{equation}
\hat B^\dag_{[4]\langle2\rangle(\tau)lm}\propto
\hat P_+\hat B^\dag_{[2]\langle2\rangle(\tau)lm},
\quad
\hat B^\dag_{[4]\langle0\rangle(0)00}\propto
\hat P_+^2.
\label{four}
\end{equation}
A four-body interaction with SO(6) PDS
is thus of the form $\hat P_+\hat V_2\hat P_-$
where $\hat V_2$ is an arbitrary two-body interaction.
This interaction leaves solvable the class of states with $\Sigma=N$
but, in general, admixes those with $\Sigma<N$.
The conclusion is that
we can construct a hierarchy of interactions
$\hat P_+^k\hat n_s\hat P_-^k,\, \hat P_+^k\hat V_2\hat P_-^k,
\ldots,\hat P_+^k\hat V_p\hat P_-^k$
of order $2k+1,\,2k+2,\ldots, 2k+p$, respectively,
that retain the SO(6)-DS in selected states.
If $\hat{V}_p$ is SO(5)-invariant,
$\tau$ is a good quantum number for all states.

\begin{figure*}
\begin{center}
\leavevmode
\includegraphics[width=0.80\linewidth]{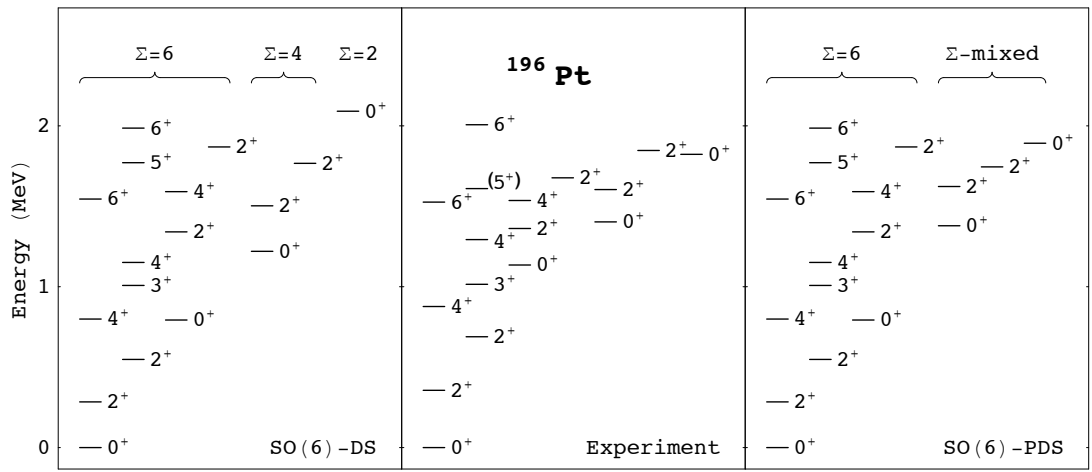}
\caption{
Observed spectrum of $^{196}$Pt~\cite{NDS07}
compared with the calculated spectra
of $\hat H_{\rm DS}$ (\ref{hDS}),
with SO(6) dynamical symmetry (DS),
and of $\hat H_{\rm PDS}$ (\ref{hPDS}) with partial dynamical symmetry (PDS).
The parameters in $\hat H_{\rm DS}$ $(\hat H_{\rm PDS})$ are
$A=174.2\, (122.9)$,
$B=44.0\, (44.0)$,
$C=17.9\, (17.9)$,
and $\eta=0\, (34.9)$ keV.
The boson number is $N=6$
and $\Sigma$ is an SO(6) label.}
\label{pt196}
\end{center}
\end{figure*}

The SO(6)-DS spectrum of Eq.~(\ref{eDS})
resembles that of a $\gamma$-unstable deformed rotor,
where states are arranged in bands with SO(6) quantum number
$\Sigma=N-2v$, $(v=0,1,2,\ldots)$.
The in-band rotational splitting
is governed by the SO(5) and SO(3) terms in $\hat{H}_{\rm DS}$ (\ref{hDS}).
A comparison with the experimental spectrum and E2 rates of $^{196}$Pt
is shown in Fig.~\ref{pt196} and Table~\ref{be2}.
\begin{table}
\caption{\label{be2}
Observed~\cite{NDS07} and calculated $B$(E2) values
(in $e^2{\rm b}^2$) for $^{196}$Pt.
For both the exact (DS) and partial (PDS)
SO(6) dynamical symmetry calculations, the E2 operator is
$e_{\rm b}[(s^\dag\times\tilde d+d^\dag\times\tilde s)^{(2)}
+\chi(d^\dag\times\tilde d)^{(2)}]$ with
$e_{\rm b}=0.151$ $e$b and $\chi=0.29$.}
\begin{ruledtabular}
\begin{tabular}{clll}
Transition& Experiment& DS &PDS\\
\hline
$2^+_1\rightarrow0^+_1$& 0.274~(1)  &  0.274&  0.274\\
$2^+_2\rightarrow2^+_1$& 0.368~(9)  &  0.358&  0.358\\
$2^+_2\rightarrow0^+_1$& 3.10$^{-8}$(3) & 0.0018& 0.0018\\
$4^+_1\rightarrow2^+_1$& 0.405~(6)  &  0.358&  0.358\\
$0^+_2\rightarrow2^+_2$& 0.121~(67) &  0.365&  0.365\\
$0^+_2\rightarrow2^+_1$& 0.019~(10) &  0.003&  0.003\\
$4^+_2\rightarrow4^+_1$& 0.115~(40) &  0.174&  0.174\\
$4^+_2\rightarrow2^+_2$& 0.196~(42) &  0.191&  0.191\\
$4^+_2\rightarrow2^+_1$& 0.004~(1)  &  0.001&  0.001\\
$6^+_1\rightarrow4^+_1$& 0.493~(32) &  0.365&  0.365\\
$2^+_3\rightarrow0^+_2$& 0.034~(34) &  0.119&  0.119\\
$2^+_3\rightarrow4^+_1$& 0.0009~(8) &  0.0004& 0.0004\\
$2^+_3\rightarrow2^+_2$& 0.0018~(16)& 0.0013& 0.0013\\
$2^+_3\rightarrow0^+_1$& 0.00002~(2)& 0        & 0         \\
$6^+_2\rightarrow6^+_1$& 0.108~(34) & 0.103& 0.103\\
$6^+_2\rightarrow4^+_2$& 0.331~(88) &  0.221&  0.221\\
$6^+_2\rightarrow4^+_1$& 0.0032~(9) & 0.0008& 0.0008\\
$0^+_3\rightarrow2^+_2$&$<0.0028$   & 0.0037& 0.0028\\
$0^+_3\rightarrow2^+_1$&$<0.034$    & 0          & 0        \\
\end{tabular}
\end{ruledtabular}
\end{table}
The SO(6)-DS limit is seen to provide a good description
for properties of states in the ground band $(\Sigma=N)$.
This observation was the basis of the claim~\cite{Cizewski78}
that the SO(6)-DS is manifested empirically in $^{196}$Pt.
However, the resulting fit to energies of excited bands is quite poor.
The $0^+_1$, $0^+_3$, and $0^+_4$ levels of $^{196}$Pt
at excitation energies 0, 1403, 1823 keV, respectively,
are identified as the bandhead states
of the ground $(v=0)$, first- $(v=1)$
and second- $(v=2)$ excited vibrational bands~\cite{Cizewski78}.
Their empirical anharmonicity,
defined by the ratio $R=E(v=2)/E(v=1)-2$,
is found to be $R=-0.70$.
In the SO(6)-DS limit these bandhead states
have $\tau=L=0$ and $\Sigma=N,N-2,N-4$, respectively.
The anharmonicity $R=-2/(N+1)$,
as calculated from Eq.~(\ref{eDS}), is fixed by $N$.
For $N=6$, which is the appropriate boson number for $^{196}$Pt,
the SO(6)-DS value is $R=-0.29$,
which is in marked disagreement with the empirical value.
A detailed study of double-phonon excitations within the IBM,
has concluded that large anharmonicities can be incorporated
only by the inclusion of at least cubic terms in the
Hamiltonian~\cite{ramos00b}.
In the IBM there are 17 possible three-body interactions.
One is thus confronted with the need to select suitable higher-order terms
that can break the DS in excited bands but preserve it in the ground band.
This can be accomplished by the PDS construction presented in this work.
On the basis of the preceding discussion
we propose to use the following Hamiltonian with SO(6)-PDS
\begin{equation}
\hat{H}_{\rm PDS}=\hat{H}_{\rm DS}+\eta\hat{P}_{+}\hat{n}_s\hat{P}_{-},
\label{hPDS}
\end{equation}
where the terms are defined in Eqs.~(\ref{hDS}) and (\ref{three}).
The spectrum of $\hat{H}_{\rm PDS}$ is shown in Fig.~\ref{pt196}.
The states belonging to the $\Sigma=N=6$ multiplet remain solvable
with energies given by the same DS expression, Eq.~(\ref{eDS}).
States with $\Sigma < 6$ are generally admixed
but agree better with the data than in the DS calculation.
For example, the bandhead states of the first- (second-) excited bands
have the SO(6) decomposition
$\Sigma=4$: $76.5\%\,(19.6\%)$,
$\Sigma=2$: $16.1\%\,(18.4\%)$,
and $\Sigma=0$: $7.4\%\,(62.0\%)$.
Thus, although the ground band is pure,
the excited bands exhibit strong SO(6) breaking.
The calculated SO(6)-PDS anharmonicity for these bands is $R=-0.63$,
much closer to the empirical value, $R=-0.70$.
We emphasize that not only the energies
but also the wave functions of the $\Sigma=N$ states remain unchanged
when the Hamiltonian is generalized from DS to PDS.
Consequently, the E2 rates for transitions among this class of states
are the same in the DS and PDS calculations.
This is evident in Table~\ref{be2}
where most of the E2 data concern transitions between $\Sigma=N=6$ states.
Only transitions involving states from excited bands ({\it e.g.},
the $0^+_3$ state in Table~\ref{be2}) can distinguish between DS and PDS.
Unfortunately, such interband E2 rates
are presently poorly known experimentally.
Their measurement is highly desirable for further
testing the SO(6)-PDS wave functions.

In conclusion, we have presented a systematic procedure
for identifying and selecting interactions, of a given order,
with partial dynamical symmetry (PDS).
This allows the construction of Hamiltonians
that break the dynamical symmetry (DS),
but retain selected subsets of solvable eigenstates with good symmetry.
As demonstrated in this work,
the advantage of using higher-order interactions with PDS
is that they can be introduced without destroying results
previously obtained with a DS for a segment of the spectrum.
These virtues generate an efficient tool
which can greatly enhance
the scope of algebraic modeling of quantum many-body systems.

This work was supported in part (A.L.) by the Israel Science
Foundation and in part (J.E.G.R.)
by the Spanish Ministerio de Educaci\'on y Ciencia, by the European
regional development fund (FEDER) under projects 
FPA2006-13807-C02-02, FPA2007-63074 and by Junta de
Andaluc\'{\i}a under projects FQM318, P07-FQM-02962.

\end{document}